\documentclass[aps,amsmath,amssymb,preprintnumbers,nofootinbib,a4paper,prl,twocolumn]{revtex4-1}
\pdfoutput=1
\usepackage{placeins}
\usepackage{xspace}
\usepackage{cancel} 
\usepackage{amssymb,url}
\usepackage{graphicx}
\usepackage{hyperref}
\usepackage{color}

\usepackage{array}
\usepackage{amsmath}
\usepackage{mathtools}
\usepackage{slashed}
\usepackage{subfigure}
\usepackage{epsfig}
\usepackage{multirow}
\usepackage[ margin=5pt, font=normalsize,labelfont=bf,justification=raggedright]{caption} 

\definecolor{rossoCP3}{cmyk}{0,.88,.77,.40}

\def\beq{\begin{equation}} 
\def\eeq{\end{equation}}

\begin{document}

\title{\Large  \color{rossoCP3} Inflation from Asymptotically Safe Theories} 
\author{Niklas Gr\o nlund Nielsen}
\email{ngnielsen@cp3-origins.net} 
\author{Francesco Sannino}
\email{sannino@cp3-origins.net} 
\author{Ole Svendsen}
\email{svendsen@cp3-origins.net} 

\affiliation{
\vspace{3mm} 
{ \color{rossoCP3}  \rm CP}$^{\color{rossoCP3} \bf 3}${\color{rossoCP3}\rm-Origins},  
University of Southern Denmark, Campusvej 55, DK-5230 Odense M, Denmark
}
 \begin{abstract}
We investigate models in which inflation is driven by an ultraviolet safe and interacting scalar sector stemming from a new class of nonsupersymmetric gauge field theories. These new theories, differently from generic scalar models, are well defined to arbitrary short distances because of the existence of a controllable ultraviolet interacting fixed point. The 
scalar couplings at the ultraviolet fixed point and their overall running are predicted by the geometric structure of the underlying theory.  We analyse the minimal and non-minimal coupling  to gravity of these theories and the consequences for inflation. In the minimal coupling case the theory requires large non-perturbative quantum corrections to the quantum potential for the theory to agree with data, while in the non-minimal coupling case the perturbative regime in the couplings of the theory is preferred. Requiring the theory to reproduce the observed amplitude of density perturbations constrain the geometric data of the theory such as the number of colors and flavors for generic values of the non-minimal coupling.   
   \\~\\[.1cm]
{\footnotesize  \it Preprint: CP$^3$-Origins-2015-008 DNRF90}
 \end{abstract}

\maketitle
 
The inflationary paradigm plays a central role in modern cosmology \cite{Guth:1980zm,Linde:1981mu}. Many realisations have appeared in the literature \cite{Martin:2014vha} with the vast majority using elementary scalar fields to drive inflation. Theories with fundamental scalars are, however, typically trivial. Meaning that for the theory to be well defined at arbitrary short scales the renormalized coupling must vanish, and consequently the resulting theory is non-interacting. It could happen that gravitational corrections can render field theories featuring scalars well defined at short distances, but so far no formal proof exists in four dimensions, and without requiring additional (space-time) symmetries. It is therefore interesting to explore models where the issue is resolved before the underlying fundamental particle theory of the inflaton is coupled to gravity. A possible solution is to assume the inflaton to be a composite state made by a more fundamental matter \cite{Channuie:2011rq,Bezrukov:2011mv} governed by an asymptotically free theory \cite{Gross:1973id,Politzer:1973fx}. The gravity dual dynamics of these models has been investigated in \cite{Anguelova:2014dza}. 
 
Recently, however, a novel class of non-trivial four-dimensional theories featuring elementary scalars appeared  \cite{Litim:2014uca}. The crucial ingredient is the presence of an exact interacting ultraviolet (UV) fixed point in all the couplings of the theory, i.e. the theories are {\it complete asymptotically safe} \cite{Litim:2014uca}.  The asymptotic safety scenario refers to the existence of high-energy fixed points \cite{WeinbergAS}. It plays a relevant role as a possible UV completion of quantum gravity  \cite{WeinbergAS,Niedermaier:2006ns,Percacci:2007sz,Litim:2011cp,Reuter:2012id}\footnote{In addition several UV conformal extensions of the standard model with(out) gravity have been discussed in literature \cite{Kazakov:2002jd,Gies:2003dp,Morris:2004mg,Fischer:2006fz,Kazakov:2007su,Zanusso:2009bs,
Gies:2009sv,Daum:2009dn,Vacca:2010mj,Calmet:2010ze,Bazzocchi:2011vr,Gies:2013pma,Antipin:2013exa,Dona:2013qba}.  
Scale invariant inspired models have also been considered in particle physics and cosmology \cite{Meissner:2006zh,Foot:2007iy,
Shaposhnikov:2009pv,Weinberg:2009wa,
Hooft:2010ac,
Hindmarsh:2011hx,Hur:2011sv,Dobrich:2012nv,Tavares:2013dga,
Tamarit:2013vda,Antipin:2013bya,
Gabrielli:2013hma,Holthausen:2013ota,Dorsch:2014qpa,Eichhorn:2014qka}.}.

The resulting physics is quite distinct from the traditional complete asymptotic freedom scenario where a non-interacting UV fixed point emerges in all the couplings \cite{Cheng:1973nv,Callaway:1988ya}; see  also \cite{Holdom:2014hla,Giudice:2014tma} for recent studies.

The template that we shall consider here consists of an $SU(N_C)$ gauge theory with $N_F$ Dirac fermions transforming according to the fundamental representation of $SU(N_C)$ and interacting with an $N_F\times N_F$ complex scalar matrix $H_{ij}$ that self-interacts. The large $N_F$ and $N_C$ Veneziano limit is taken such that the ratio $N_F/N_C$ is a continuous parameter. The details of the model are given in \cite{Litim:2014uca}. It is useful to introduce the positive control parameter $\delta = N_F/N_C - 11/2$  that can be taken to be arbitrarily small\footnote{$\delta$ corresponds to $\epsilon$ in  \cite{Litim:2014uca}. We switched to $\delta$  to  avoid misunderstanding with respect to the standard notation for one of the slow-roll parameters.}.  The hypercritical surface, in the four-dimensional coupling space, is unidimensional. This implies that along the {\it line of physics}, which is the globally defined renormalization group line connecting the infrared and the UV fixed point, the dynamics is driven by a single coupling, e.g. the gauge coupling. All the other couplings, including the scalar ones, follow the gauge one. Furthermore in \cite{Litim:2015iea} it has been shown that the scalar potential is stable at the classical and quantum level.  Therefore these theories hold a special status, they are fundamental according to Wilson's definition and we shall use it to model inflation.  The first phenomenological application of these kind of theories appeared in \cite{Sannino:2014lxa}. 

The quantum corrected and leading-log resummed potential along the RG flow from the infrared to the ultraviolet reads \cite{Litim:2015iea}:
\beq\label{V}
\begin{array}{rcl}
V_{\rm iUVFP}(\phi)&=& 
\displaystyle
\frac{\lambda_*\, \phi^4}{4N_f^2(1+W(\phi))}\left(\frac{W(\phi)}{W(\mu_0)}\right)^{\frac{18}{13\delta}} \ ,
\end{array}
\eeq
where the positive quartic coupling is given by $\lambda_*=\delta\,\frac{16\pi^2}{19}(\sqrt{20+6\sqrt{23}}-\sqrt{23}-1)$ at the fixed point. $\phi$ is the real scalar field along the diagonal of $H_{ij} = \phi\, \delta_{ij}/\sqrt{2N_f}$. The different normalisation in $H_{ij}$ with respect to Ref. \cite{Litim:2015iea} ensures a canonically normalised kinetic term for $\phi$.  $W(\phi)\equiv W[z(\phi)]$, where $W[z]$ is the Lambert function solving the transcendent equation
\begin{eqnarray}
z &=& W \exp W  \ , \quad {\rm with}\\ 
z(\mu)&=&\left(\frac{\mu_0}{\mu}\right)^{\frac{4}{3}\delta\alpha^*} \left(\frac{\alpha^*}{\alpha_0}-1\right)\exp\left[\frac{\alpha^*}{\alpha_0}-1\right]  \ .\label{eqn:z}
\end{eqnarray}
 $\alpha^* = \frac{26}{57} \delta + {\cal O} (\delta^2)$ is the gauge coupling at its UV fixed point value and $\alpha_0=\alpha(\mu_0)$ the same coupling at a reference scale $\mu_0$. The asymptotically safe nature is easily grasped by showing the explicit running of the coupling: 
 \begin{equation}
 \alpha = \frac{\alpha^{\ast}}{1 + W(\mu)} \ . \label{eq:gauge coupling}
 \end{equation}
 At asymptotically high energies $W(\mu)$ vanishes while it grows towards the infrared. It is convenient to fix $\alpha_0$ via $\alpha_0 \equiv \alpha^*/(1+k)$ with $k \in \mathbb{R}_+$ which, in practice, amounts to fix the arbitrary renormalization reference scale $\mu_0$  along the RG flow. As pointed out in \cite{Litim:2015iea} the value of $k=1/2$, i.e. $\alpha_0 = 2\alpha^*/3$,   corresponds to an exact critical transition scale $\mu_0 = \Lambda_c$  above which the physics is dominated by the interacting UV fixed point and below it by the gaussian IR fixed point.  The interacting nature of the UV fixed point embodies the fact that it is approached as a power law in the renormalization scale 
 \begin{equation}
\alpha (\mu ) = \alpha^\ast + (\alpha (\mu_0) - \alpha^\ast) \left( \frac{\mu}{\mu_0}\right)^{-\tfrac{104}{171} \delta^2 + \mathcal{O}(\delta^3)},
\end{equation}
   along the  line of physics.

 The Lambert function  in the deep UV limit  approaches
\begin{equation}
\lim_{\phi/\mu_0 \to \infty}  W(\phi) = k\left( \frac{\phi}{\mu_0}\right)^{-\tfrac{104}{171} \delta^2}.
\end{equation}
Here we have replaced the renormalisation scale with the value of the background scalar field value $\phi$. 
The potential therefore acquires the asymptotic form  
\begin{equation}
\lim_{\phi/\mu_0 \to \infty} V_{\rm iUVFP} = \frac{\lambda_{\ast} \phi^4}{4N_F^2} \left( \frac{\phi}{\mu_0}\right)^{-\tfrac{16}{19} \delta}.
\label{eq:UVpotential}
\end{equation}
 Because $\delta > 0$ the overall exponent is reduced, at high energies, with respect to the classical theory. In a theory with an interacting UV fixed point we observe that the overall coupling and exponent are geometric quantities that depend solely on the number of flavours and colours of the theory. Furthermore the overall height of the potential can be made arbitrary small by reducing $\delta$, which, de facto leads to a small amplitude of scalar perturbations as will be shown. 
 
 The template offers the opportunity to investigate the inflationary dynamics of asymptotically safe gauge theories and to grasp some of its general features.

  \section{Gravity and Inflation}
We  couple the model to gravity as follows \begin{equation}
S_{\rm J}= \int d^4x \sqrt{-g} \left\{ - \frac{M^2+\xi \phi^2}{2}R + \frac{g^{\mu \nu}}{2} \partial_\mu \phi\partial_\nu \phi - V_{\rm iUVFP} \right\} \ ,  
\end{equation}
where, for simplicity, we only show the coupling to $\phi$, the modulus of $H$, that we take to drive inflation. 
A conformal transformation of the metric allows to rewrite the action as minimally coupled but with a new canonically normalised scalar field and potential.  This will transform to the so called Einstein frame from the original Jordan frame \cite{Joergensen:2014rya,Bezrukov:2011mv}.
 
 We will now examine the inflationary predictions of this potential, assuming single field slow roll inflation, by first computing the associated slow roll parameters  \cite{Ade:2015lrj}
\begin{equation}
\epsilon = \frac{M_{\rm P}^2}{2} \left( \frac{dU/d\chi}{U} \right)^2 \ ,
\end{equation}
and
\begin{equation}
\eta = M_{\rm P}^2 \frac{d^2U/d\chi^2}{U} \label{secsr}\ .
\end{equation}
Here $U=V_{\rm iUVFP}/\Omega^4$ with $\Omega^2=(M^2+\xi\phi^2)/M_{\rm P}^2$ the conformal transformation of the metric and  $\chi$ the canonically normalized field in the Einstein frame. 
Note that throughout this paper we will assume $M=M_{\rm P}$. In the future it would be interesting to analyse also the induced gravity limit  \cite{Accetta:1985du,CervantesCota:1995tz,Fakir:1990iu,Kallosh:2013maa,Kallosh:2013hoa}.

Inflation ends when the slow roll conditions are violated, that is when $\epsilon(\phi_{\rm end}) = 1$ or $|\eta(\phi_{\rm end} )|= 1$. The number of $e$-folds is
\begin{equation}
\label{defefold}
N = \frac{1}{M_{\rm P}^2} \int_{\chi_{\rm  end}}^{\chi_{\rm ini}}\frac{U}{dU/d\chi}d\chi\, ,
\end{equation}
which we will set to $N = 60$. 
In this work we will compare to the experimental results via the power spectrum of scalar perturbation of CMB, namely the amplitude $A_s$ and tilt $n_s$, and the relative strength of tensor perturbations, i.e. the tensor-to-scalar ratio $r$. In terms of slow roll parameters these are given by
\begin{eqnarray}
A_s &=& \frac{U}{24\pi^2 M_P^4 \epsilon} \ ,\label{eq:amplitude}\\
n_s &=&1+2\eta-6\epsilon, \quad \label{eq:tilt}\\
r&=&16\epsilon \label{eq:ttsr}\ , 
\end{eqnarray}
where all parameters are evaluated at the field value $\chi_{\rm in}$.

The analysis will be made independently for the minimally ($\xi =0$) and for the non-minimally coupled scenario ($\xi > 0$).

\section{ Minimal coupling  }
The inflationary potential here is directly $ V_{\rm iUVFP}$. For each given value of $\phi$ the overall height of the potential decreases with increasing $N_F$, decreasing $\delta$, and/or by decreasing the reference scale $\mu_0$ above which the UV fixed point is nearly reached. It is therefore clear that the model allows for several natural ways to achieve the observed amplitude of scalar perturbations. The naturality resides in the fact that these parameters that we are allowed to change are all geometric in nature, i.e. depend on the structural properties of the underlying theory like the number of flavors and colors. This is different from the usual inflationary single-field paradigm where a scalar self-coupling, a priori of order one, must be fine-tuned to a tiny value. 

\begin{figure}
\centering
\includegraphics[width=.35\textwidth]{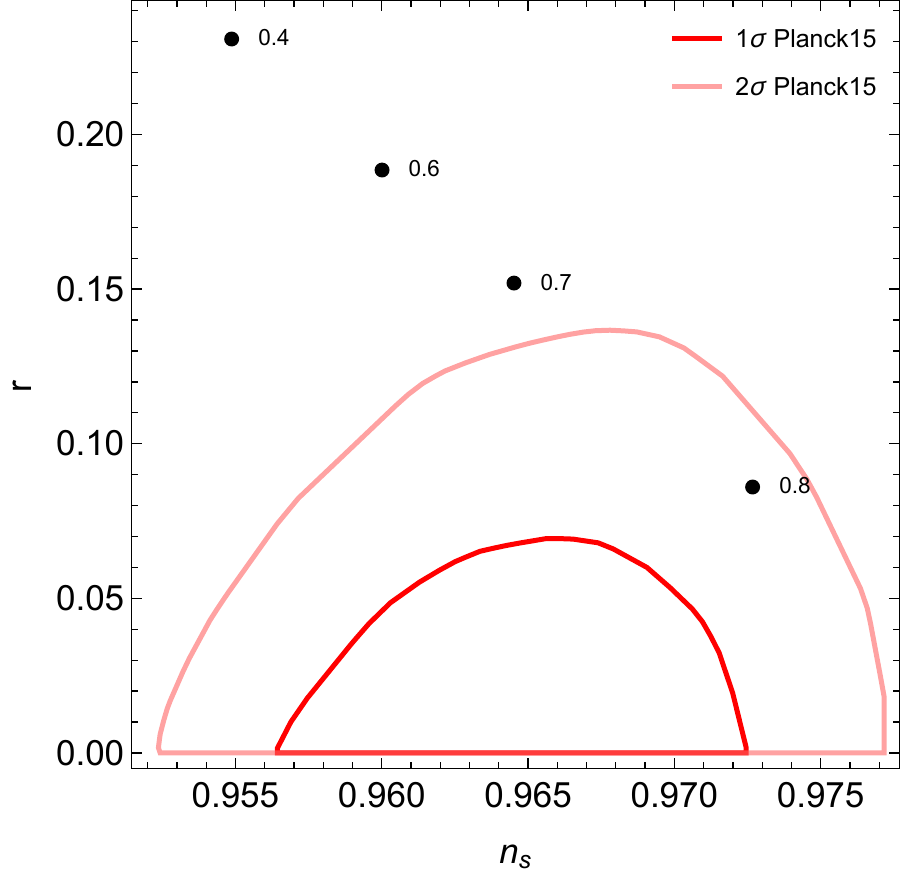}
\vspace{-10pt}
\caption{ We compare the theoretical predictions in the $r$ -$n_s$ plane for different values of $\delta$ with Planck '15 results for TT, TE, EE, +lowP and assuming $\Lambda$CDM + r \cite{Ade:2015lrj}. We used the complete expression for the quantum corrected potential in \eqref{V} and further  assumed $\mu_0 = 10^{-3} M_{\rm P}$}
\label{fig:minimalrnsPlanck}
\vspace{-10pt}
\end{figure}

\begin{figure}
\centering
\includegraphics[width=.35\textwidth]{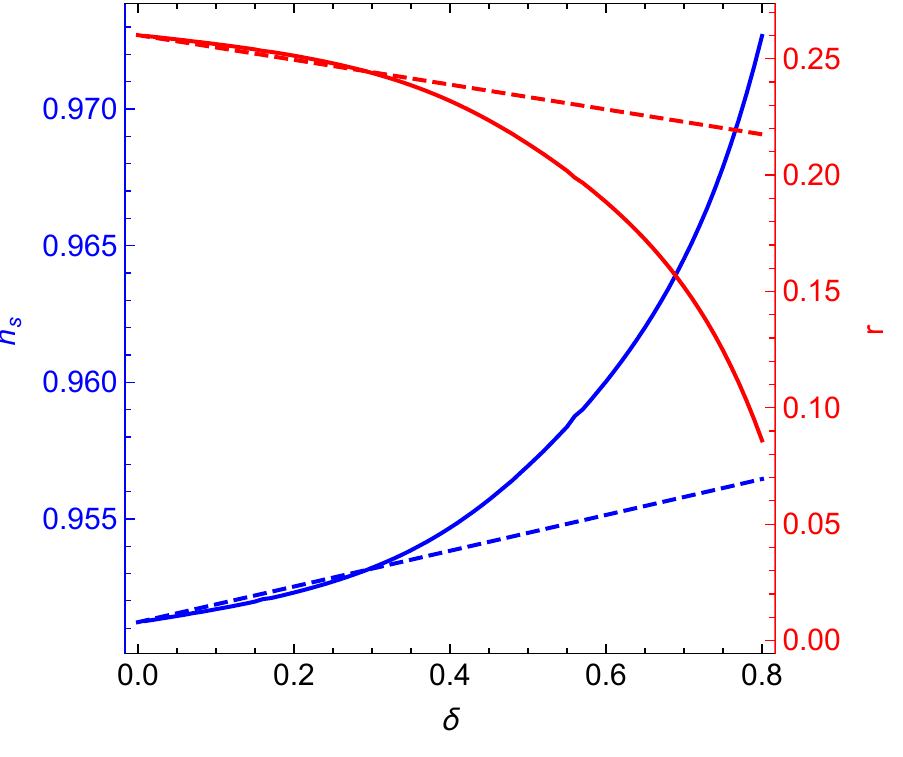}
\vspace{-15pt}
\caption{ This figure shows $r$ in red (upper curves at small $\delta$) and $n_s$ in blue as function of $\delta$. The solid lines are calculated using the complete expression for the potential in  \eqref{V} and further assuming $\mu_0 = 10^{-3} M_{\rm P}$. The dashed lines show the leading order in $\delta$ from equations \eqref{eq:ns_est} and \eqref{eq:r_est}.  }
\label{fig:full_vs_UVrns}
\vspace{-20pt}
\end{figure}

For the UV potential \eqref{eq:UVpotential} the field value at the end of inflation $\phi_{\rm end}$ reads
 \begin{equation}
\phi_{\rm end}=\sqrt{(4-\tfrac{16}{19} \delta)(3-\tfrac{16}{19} \delta)} \;M_{\rm P}
\end{equation}
The initial value of the field for $N$ $e$-folds reads 
\begin{equation}
\phi_{\rm in} = \sqrt{(4-\tfrac{16}{19} \delta)(2N + 3-\tfrac{16}{19} \delta)}\; M_{\rm P}.
\end{equation}
We observe that the corrections to the anomalous dimension of the scalar field, parametrised by $\delta $, tends to lower the field values of inflation that, however, remain transplanckian. For $N=60$ $e$-folds we have for $r$ and $n_s$\footnote{These results correct the ones in Equation (33) and (34) of \cite{Joergensen:2014rya} because it is $\eta$ that violates first the slow roll condition and not $\epsilon$, as it was assumed in \cite{Joergensen:2014rya}.}

\begin{equation}
n_s =  \frac{2N - 3}{ 2 N+3 -\tfrac{16}{19} \delta} = 0.951+0.00651 \cdot \delta + \mathcal{O}(\delta^2),  \label{eq:ns_est}
\end{equation}
\begin{equation}
r=
\frac{32(1-\tfrac{4}{19}\delta)}{2N+3-\tfrac{16}{19} \delta} = 0.260-0.0530 \cdot \delta+ \mathcal{O}(\delta^2). \label{eq:r_est}
\end{equation}
 These results are shown in show in  Fig.~\ref{fig:minimalrnsPlanck}. This shows that to be within the $2 \sigma$ Planck '15  contours  \cite{Ade:2015lrj}  values of  $\delta$  around $0.7 - 0.8$ are needed.  These relatively large values are outside the perturbative regime of the theory. The importance of higher order corrections can be deduced from Fig.~\ref{fig:full_vs_UVrns} where we show the comparison of the linear approximation in $\delta$ with  the full dependence stemming from the potential in equation \eqref{V}. 

Using  \eqref{eq:UVpotential} and \eqref{eq:amplitude} we compute the amplitude of scalar perturbations 
\begin{align}
A_s = \frac{\lambda_\ast}{48 \pi^2(4-\tfrac{16}{19}\delta)^2 N_F^2} & \left( \frac{\phi_{\rm in}}{M_{\rm P}}\right)^{6-\tfrac{16}{19}\delta}\left( \frac{\mu_0}{M_{\rm P}}\right)^{\tfrac{16}{19}\delta}\\
&\sim \frac{10^5 \cdot \delta}{N_F^2}\left( \frac{\mu_0}{M_{\rm P}}\right)^{\tfrac{16}{19}\delta} \notag.
\end{align}
Requiring $A_s = 2.2 \cdot 10^{-9}$, as measured by Planck '15, allows to determine the following relationship between the transition scale $\mu_0$, $\delta$ and $N_F$ 
\begin{equation}
  2 \log_{10} N_F -\log_{10} \delta -\frac{16 \delta}{19}\log_{10} \left(\frac{\mu_0}{M_{\rm P}} \right) \approx 14 . \label{eq:orderofMagEST}
\end{equation}
If we, for example, require the transition scale to be close to the  grand-unified inspired energy scale $\sim 10^{-3} M_{\rm P}$ and further assume $\delta = 0.1$ we obtain $N_F \sim 10^6$.  The needed number of flavors drops quickly when $\delta$ increases towards the values preferred by the Planck results.

\section{Non-minimal coupling }
\begin{figure}
\centering
\includegraphics[width=.35\textwidth]{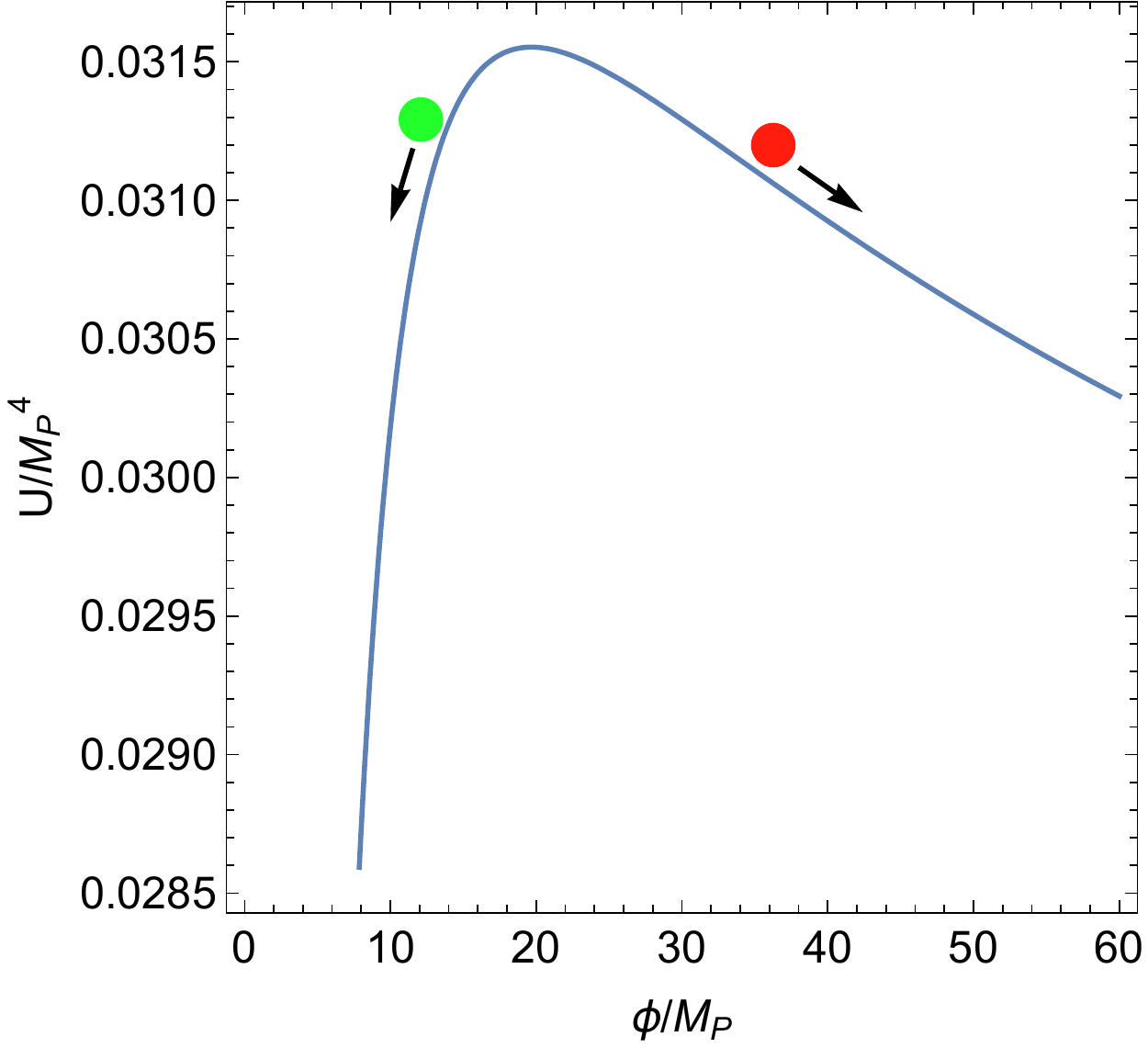}
\vspace{-8pt}
\caption{The non-minimally coupled potential for $\delta = 0.1$, $N_F = 10$, $\xi=1/6$, $\mu_0 = 10^{-3} M_{\rm P}$.   }
\label{fig:Nonminimalpot}
\end{figure}

Here the non-minimal coupling parameter $\xi$ is non vanishing. The Einstein frame potential is 
\begin{equation}
U  = \frac{V_{\rm iUVFP}}{\Omega^4} \approx \frac{\lambda_\ast \phi^4}{4N_F^2\left(1+\tfrac{\xi \phi^2}{M_{\rm P}^2}\right)^2} \left(\frac{\phi}{\mu_0} \right)^{-\tfrac{16}{19}\delta}.\label{eq:NonminPot}
\end{equation}
We plot the potential in Fig.~\ref{fig:Nonminimalpot}.  In the large field limit $\phi\gg M_{\rm P}/\sqrt{\xi}$ the $\phi^4$ term in the numerator cancels against the term in the denominator. In this limit the quantum corrections dictate the behaviour of the potential, which is found to decrease as:
\begin{equation}
\frac{\lambda_\ast M_{\rm P}^4}{4N_F^2 \xi^2} \left(\frac{\phi}{\mu_0} \right)^{-\tfrac{16}{19}\delta}.
\end{equation}
This is the region of the potential to the right of the maximum in Fig.~\ref{fig:Nonminimalpot}. Inflation could, in principle, occur on this side of the potential, naively indicated by the rolling of the red ball.  However  since the potential flattens out with increasing $\phi$ this option is not viable because the theory, in isolation, does not permit a violation of the slow roll conditions. This would, in fact, lead to a never ending slowly rolling  inflationary epoch. 

\begin{figure}
\centering
\includegraphics[width=.35\textwidth]{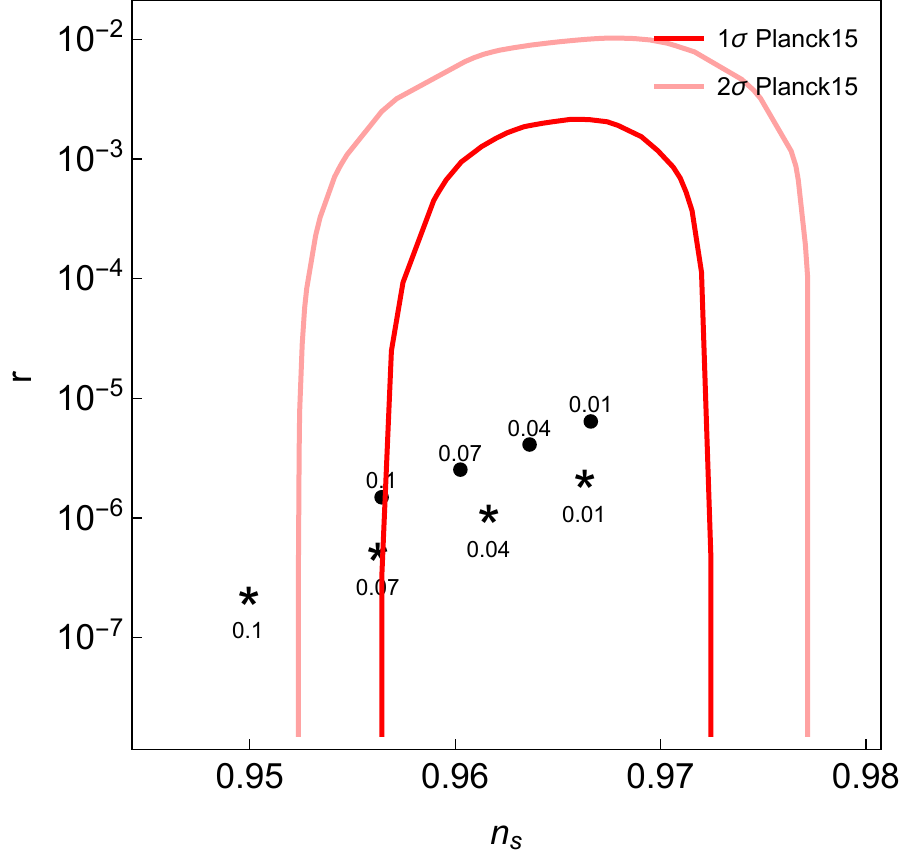}
\vspace{-10pt}
\caption{We compare the theoretical predictions in the $r$ -$n_s$ plane, in the non-minimally coupled case, for different values of $\delta$ with Planck '15 results \cite{Ade:2015lrj}. Full dots refer to the conformal coupling choice for $\xi = 1/6$ and the $\ast$ marked points to $\xi = 10^{3}$.  We used the complete expression for the quantum corrected potential in \eqref{V} and further  assumed $\mu_0 = 10^{-3} M_{\rm P}$ and the number of $e$-folds is 60.}
\label{fig:NonminimalRNs}
\end{figure}

We will therefore concentrate on the region to the left of the maximum, indicated by the green rolling ball, where it is seen that inflation can be brought to an end. 
 Furthermore the resulting $r$ and $n_s$ values agree with the Planck '15 measurements. We show in Fig.~\ref{fig:NonminimalRNs}  the $r$ and $n_s$ predictions for different $\delta$ for 60 $e$-folds and for either $\xi = 1/6$ or $\xi =10^3$. In agreement with attractor-type models \cite{Kallosh:2013tua,Kallosh:2013maa,Codello:2014sua,Joergensen:2014rya} the tensor-to-scalar ratio is small and the $n_s$ predictions are mostly inside the Planck contours. The larger is $\xi$ and the more the results are sensitive to  increasing $\delta$. Differently from the non-minimal coupling case we are well within the Planck allowed regions for values of $\delta$ compatible with perturbation theory of the underlying fundamental inflationary dynamics.  
 
\begin{figure}
\centering
\includegraphics[width=.35\textwidth]{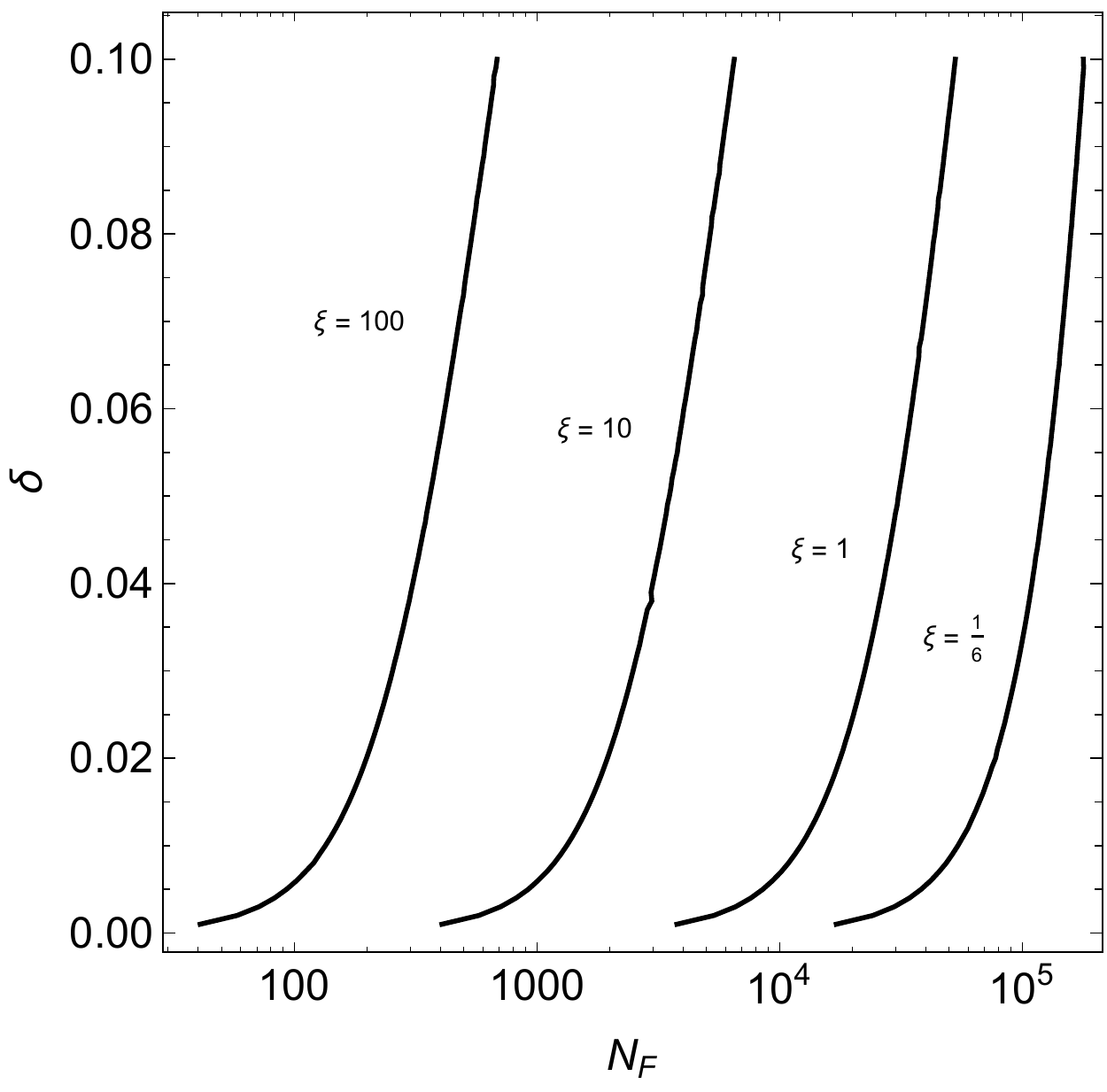}
\vspace{-10pt}
\caption{This figure shows the $\delta$ dependence on $N_F$ for different values of the non-minimal coupling $\xi$ obtained by constraining the model to provide the observed amplitude of density perturbations. The plot assumes the transition scale $\mu_0 = 10^{-3} M_{\rm P}$.}
\label{fig:Nfvsdeltaxi}
\end{figure}

Requiring the theory to produce the correct value of the amplitude of density perturbations relates the $\mu_0$, $\delta$, $N_F$ and now also the $\xi$ parameters. We show in Fig.~ \ref{fig:Nfvsdeltaxi} the resulting dependence of $\delta$ on $N_f$ for fixed  $\mu_0=10^{-3} M_{\rm P}$ and several values of $\xi$. There is also a rather weak dependence on the choice of $\mu_0/M_{\rm P}$, since it enters the potential with a power of $(16/19)\delta$. $N_F$ decreases fast with increasing $\xi$ but also with decreasing $\delta$ because of the further underlying theory relation $\lambda_\ast \propto \delta$.

\begin{figure}
\centering
\includegraphics[width=.35\textwidth]{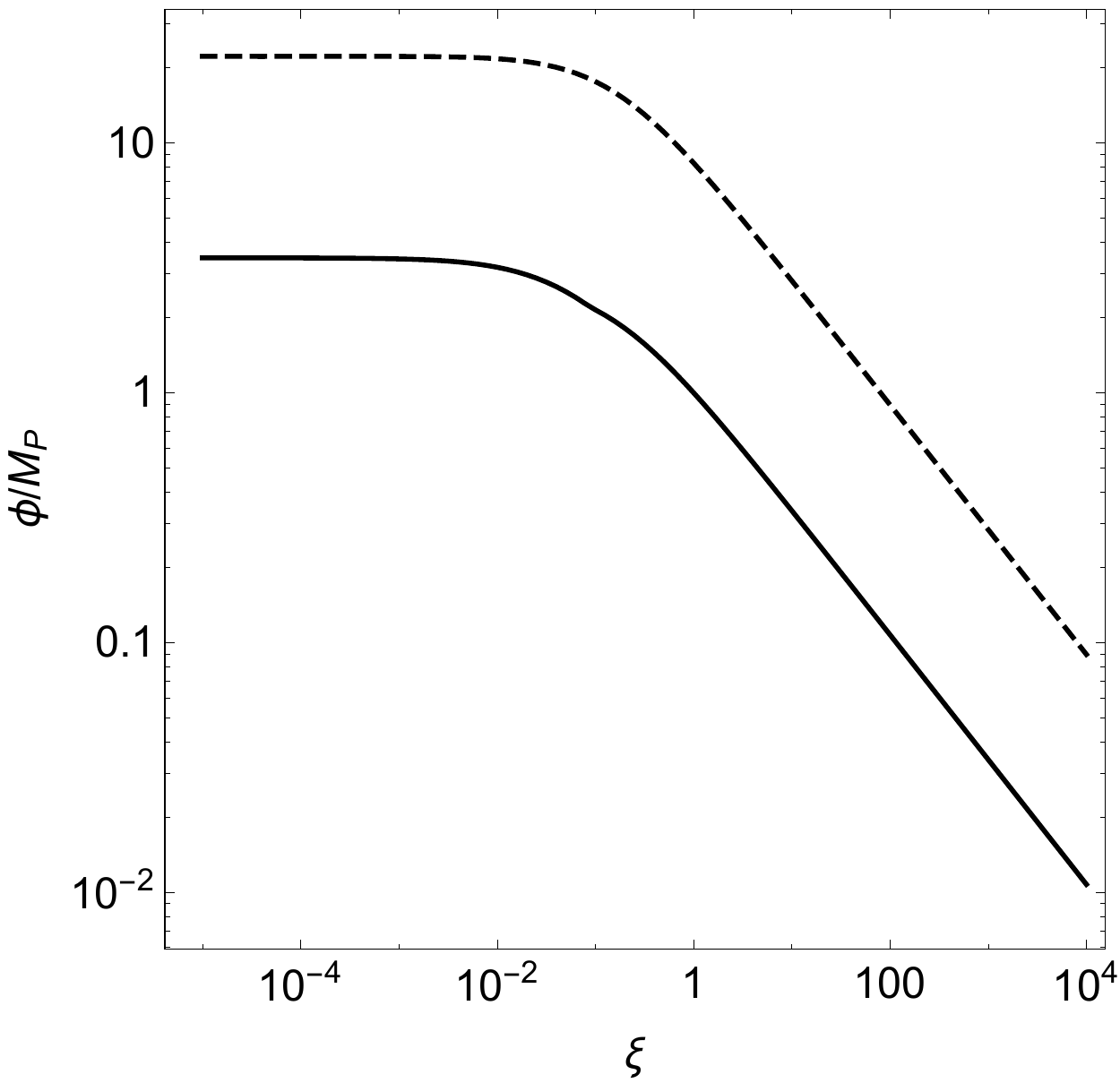}
\vspace{-10pt}
\caption{The figure shows the initial (dashed-line) and final (solid-line) values of the inflaton field in the Jordan frame as function of the non-minimal coupling $\xi$ for $\delta = 0.01$ and $\mu_0 = 10^{-3} M_{\rm P}$.}
\label{fields}
\end{figure}

For completeness we show in Fig.~\ref{fields} the initial (dashed-line) and final (solid-line) values of the field in units of the Planck scale as function of the non-minimal coupling $\xi$. The figure demonstrates that these values decrease below the Planck scale for $\xi$ above the conformal value and approach constant transplankian values above the Planck scale for small $\xi$. 

\section{Conclusion and Self Criticism}

We introduced models of inflation stemming from complete asymptotically safe field theories. The novelty resides in the fact that, differently from generic scalar models, the theories, before coupling to gravity and without additional symmetries such as supersymmetry or extra space-time dimensions, are well defined  to arbitrary short distances because of the existence of the controllable UV interacting fixed point.  The 
scalar couplings and their  running are predicted by the geometric structure of the underlying theory \cite{Litim:2014uca}.  The quantum potential has been computed in \cite{Litim:2015iea}. We could therefore use it to analyse the minimal and non-minimal coupling to gravity and its consequences for inflation. We have shown that inflation can occur in both cases. The minimal coupling case requires large non-perturbative corrections to the potential for the theory to agree with data, while the non-minimal coupling prefers the perturbative regime of the theory. Furthermore the observed value of the amplitude of density perturbations helps selecting the geometric data of the theory, i.e. the number of colors and flavors, for generic values of the non-minimal coupling. In particular one can achieve a successful inflationary scenario even for $\xi=1/6$, i.e. the conformal value. 

Despite these partial successes we still face several challenges. Gravity, for example, in our investigation has played a spectator role. It is conceivable that once its dynamics is taken into consideration, in a controllable manner, it might modify the UV behaviour of the theory.  In this case one can imagine the possible existence of a combined UV interacting fixed point of the resulting theory. If gravity itself develops an UV interacting fixed point as suggested by Weinberg  \cite{Weinberg:2009wa} it might also drive inflation \cite{WeinbergAS}.   In the future it would be interesting to analyse or perhaps even unify these, so far, complementary avenues. 

 \vskip .2cm
 \noindent
We thank Daniel Litim for relevant discussions. The work is partially supported by the Danish National Research Foundation under the grant number DNRF:90.

\end{document}